# Observation of Phonon Angular Momentum

H. Zhang[1], N. Peshcherenko[2], Fazhi Yang[1], T. Z. Ward[1], P. Raghuvanshi[1], L. Lindsay[1], Claudia Felser[2], Y. Zhang[3,4], J.-Q. Yan[1], H. Miao[1]

[1]*Materials Science and Technology Division, Oak Ridge National Laboratory, Oak Ridge, Tennessee 37831, USA*

[2]*Max Planck Institute for Chemical Physics of Solids, Nöthnitzer Strasse 40, Dresden, Germany.*

[3]*Department of Physics and Astronomy, The University of Tennessee, Knoxville, Tennessee 37996, USA*

[4]*Min H. Kao Department of Electrical Engineering and Computer Science, University of Tennessee, Knoxville, Tennessee 37996, USA*

**Angular momentum (AM), a fundamental concept describing the rotation of an object about an axis, profoundly influences all branches of physics. In condensed matter, AM is intimately related to the emergence of topological quantum states, including chiral superconductivity and quantum spin liquids[1], and various chiral quasiparticles[2-4]. Recently, it has been predicted that microscopic lattice excitations, known as phonons, can carry finite AM[4] with remarkable macroscopic physical consequences[5-11]. However, the direct observation of phonon-AM has not been achieved. In this letter, we report the experimental discovery of phonon-AM in the chiral crystal Tellurium. We show that due to AM conservation, applying a time-reversal symmetry breaking thermal gradient along the chiral axis of single crystal Te results in a macroscopic mechanical torque, $\tau$, that can be observed using a cantilever-based device. We establish that the mechanical torques change sign by flipping the thermal gradient and disappear in polycrystalline samples that lack a preferred chirality. Based on our experimental settings, we estimate $\tau \sim 10^{-11}$ N · m, in agreement with theoretical calculations. Our results uncover phonon-AM and pave the way for phonon-AM enabled quantum states for microelectronic applications.**

In 1915, Einstein and de Haas discovered mechanical rotations of a suspended ferromagnetic cylinder under external magnetic field (Fig. 1**a**)[12]. This effect, now known as the Einstein-de Haas effect (EdH), connects quantum mechanical spin and classical angular momentum (AM) and is regarded as one of the milestones in the history of quantum mechanics[13]. Recently, in analogy to the EdH for the spin-AM, a phonon version of the EdH has been theoretically proposed for non-centrosymmetric materials[5], such as chiral elemental Tellurium (Te) single crystal, which hosts phonon modes characterized by circular atomic motions (Fig. 1**b** and **c**). As schematically shown in Fig. 1**d**, applying a thermal gradient, $\nabla T$, breaks the time-reversal symmetry, $\mathcal{T}$, of the chiral crystal and can induce finite phonon-AM. Due to AM conservation, mechanical rotations of the chiral crystal are required to compensate the thermal gradient induced internal phonon-AM. The phonon EdH effect represents a macroscopic manifestation of the microscopic chiral quantum excitations of the lattice with profound fundamental implications and potential for applications such as dark matter detection, giant phonon magnetic moment, and quantum transductions[5-11]. In this letter, we demonstrate phonon-AM and subsequent phonon EdH in a prototypical non-magnetic chiral crystal Te using a cantilever-based device.

In equilibrium, the phonon-AM is expressed as[4,5]:

$$\boldsymbol{J}_{ph} = \sum_{\boldsymbol{k},\sigma} \boldsymbol{l}_\sigma(\boldsymbol{k}) \left[ f_0(\omega_\sigma(\boldsymbol{k})) + \frac{1}{2} \right], \quad (1)$$

where $\boldsymbol{l}_\sigma(\boldsymbol{k}) = \hbar \boldsymbol{\epsilon}_\sigma^\dagger(\boldsymbol{k}) \boldsymbol{M} \boldsymbol{\epsilon}_\sigma(\boldsymbol{k})$ and $\omega_\sigma(\boldsymbol{k})$ are the AM and frequency of phonon mode, $\sigma$, with momentum, $\boldsymbol{k}$. $f_0(\omega_\sigma(\boldsymbol{k})) = 1/(e^{\frac{\hbar \omega_\sigma(\boldsymbol{k})}{k_B T}} - 1)$ is the Bose distribution and $\boldsymbol{\epsilon}_\sigma(\boldsymbol{k})$ is the normalized eigenvector. The *N*-by-*N* unit matrix $\boldsymbol{M}$ describes the 3-dimensional rotation in a unit cell with $N$ atoms (see Methods). In the presence of $\mathcal{T}$-symmetry, $\boldsymbol{l}_\sigma(\boldsymbol{k}) = -\boldsymbol{l}_\sigma(-\boldsymbol{k})$ ensures $\boldsymbol{J}_{ph} = 0$ in equilibrium. Applying $\nabla T$ along the chiral crystal drives the phonon system from equilibrium and breaks $\mathcal{T}$ thus allowing for finite $\boldsymbol{J}_{ph}$ below the characteristic Debye temperature scale, $\theta_D$.[4]

The classic EdH experiment, depicted in Fig. 1, requires a ferromagnet to rotate with minimum friction in order to measure its rotation, which implies that the crystal rod cannot be firmly held to either base. This requirement is, however, challenging for the phonon-EdH as building $\nabla T$ in a cryogenic environment requires solid physical contacts for thermalization and flexibility for distortion monitoring. This challenge calls for an innovation in experimental design. Fundamentally, the phonon-AM will also generate a mechanical torque, $\boldsymbol{\tau} = \frac{J_{ph}}{t_0}$, where $t_0$ is the overall phonon relaxation time. Using modern micro-cantilever technologies[14], this torque can then be measured with high accuracy. Figure 1**e** depicts our experimental setup for the torque measurement: chiral Te crystals are connected to cantilevers that are coupled to cooling reservoirs (See Methods). A thermal gradient is created by shining a laser on the sample. The direction of thermal energy flow can be controlled by moving the laser spot on the sample. The phonon-AM induced torque bends the cantilevers depending on the direction of the induced $\tau$ as shown in the insets of Fig. 1e.

Figures 2a and 2b show pictures of the real experimental setups for single crystal and polycrystalline Te samples, respectively. To distinguish the torque induced by the intrinsic phonon-AM from the effects of thermal expansion (see Methods), we designed the cantilevers to be aligned antiparallelly. In this geometry, the phonon-AM induced $\tau$ is expected to be opposite between the two cantilevers, while the thermal expansion will have the same sign for each. Figure 2c shows the torque responses of single crystal Te at $T=10$ K. The 25-mW and 532 nm laser is periodically turned on and off while the torque responses are reordered in a time interval (between adjacent data points in Fig. 2c) of 5 seconds. Opposite $\tau$ on the order of $10^{-11}$ N·m is observed for the separate cantilevers with laser heating. $\tau$ for both cantilevers rapidly drops back to zero when the laser heating is turned off. In stark contrast, in the control experiment using a polycrystalline Te sample with no net chiral phonon heat flow, both cantilevers display the same negative sign of $\tau$ with the magnitude decreasing monotonically at elevated temperature, consistent with the effect of heating. This control experiment establishes a chirality induced $\tau$ in the Te single crystal.

Another key signature of phonon-AM is the change in sign when the thermal gradient is reversed, which should manifest in a change in the sign of our measured torques. To switch the thermal gradient, we move the laser spot to different sites on the samples. The laser is turned off before moving to different positions. Figure 3a shows the laser position dependence of $\tau$. Remarkably, at $T = 10$ K, the torques on both cantilevers switch signs as the laser spot moves from Cantilever 2 to Cantilever 1, firmly establishing the opposite torque under opposite directions of applied thermal gradient. We applied the same procedure on the polycrystalline crystal at 10 K (Fig. 3c) and find that the signs of $\tau$ on both cantilevers remain unchanged for all laser position. Here, the heating effect is more obviously revealed where larger $\tau$ is observed on the cantilever that is closer the laser spot. It has been shown theoretically that the phonon-AM approaches to zero for $T \gg \theta_D$[4]. For Te single crystals, $\theta_D \sim 130$ K[15]. To verify these predictions, we repeat the same experiment shown in Fig. 3a at room temperature, $\sim 2.3 \times \theta_D$ of Te. As shown in Fig. 3b, the torques from both cantilevers display the same sign, in agreement with the significantly reduced phonon-AM at high $T$.

Finally, we compare the magnitudes of the experimentally determined $\tau$ with theoretical estimations. As mentioned above, $\boldsymbol{J}_{ph}$ arises from chiral phonon distributions deviating from Bose-Einstein statistics due to the applied temperature gradient, which is fundamentally similar to the Edelstein effect in electronic systems[5,16]. Based on the Boltzmann equation and the relaxation time approximation, the phonon distribution function, $n_\sigma(\boldsymbol{q})$, under an applied thermal gradient is:

$$n_\sigma(\boldsymbol{k}) = f_0(\omega_\sigma(\boldsymbol{k})) - t_0(\boldsymbol{v}_\sigma(\boldsymbol{k}) \cdot \nabla T)\partial_T f_0(\omega_\sigma(\boldsymbol{k})). \quad (2)$$

where $\boldsymbol{v}_\sigma(\boldsymbol{k})$ is the group velocity of the phonon mode $\sigma$. Following the angular momentum density of Eq. (1), it then predicts:

$$\delta J_{ph} = \int \frac{d^3k}{(2\pi)^3} \sum_\sigma l_\sigma(\boldsymbol{k}) \tau (v_\sigma(\boldsymbol{k}) \cdot \nabla T) \frac{\omega_\sigma(\boldsymbol{k})}{T} \frac{\partial n_{eq}(\omega_\sigma(\boldsymbol{k}))}{\partial \omega_\sigma(\boldsymbol{k})}, \quad (4)$$

where the integration runs over phonon momenta $\boldsymbol{k}$ sampled in the first Brillouin zone. In the relevant temperature domain ($T = 10\ K \ll \theta_D$), the angular momentum density $\delta J_{ph}$ is dominated by phonon contributions with energies $\hbar\omega_\sigma(\boldsymbol{k}) \sim k_B T$. Assuming that the speed of sound of the Te single crystal is $v \sim 10^3 m \cdot s^{-1}$ and phonon relaxation time $t_0 \sim 10^{-8}\ s$,[5] a simple estimation of Eq. (2) for angular momentum density per $\nabla T$ gives $\delta J_{ph} \sim 10^{-13} J \cdot s/(m^2 \cdot K)$, which predicts the total exerted mechanical torque, $\tau \sim \frac{\delta J_{ph}}{t_0} V$, where $V$ is sample volume. Putting the experimental sample dimensions $V = 1 \times 0.2 \times 0.1$ mm³ and assuming a plausible temperature gradient $\nabla T = 10$ K/mm, the calculated mechanical torque is $\tau \sim 10^{-12}\ N \cdot m$, in reasonable agreement with the experimental findings of $\tau \sim 10^{-11} N \cdot m$.

In summary, our measurements firmly establish the existence of phonon-AM in chiral crystals. Phonon-AM is the theoretical basis of chiral and topological phonons[11] that may interact with topological fermions to create unique topological quantum states[7,17-19]. Phonons can also transfer AM to other fundamental particles and elementary excitations allowing for novel quantum transduction mechanisms[5,9,10], thermal manipulation of spin,[8,20,21] and detection of hidden quantum fields[6,22,23]. This discovery provides a solid foundation for emergent chiral quantum states and opens a new avenue for phonon-AM enabled quantum information science and microelectronic applications.

# Methods

**Crystal synthesis:**

Tellurium single crystals were grown by the physical vapor transport method[24]. Tellurium powders were sealed in a quartz ampule and put into a two-zone furnace. The hot and cold zones were kept at 450 °C and 360 °C, respectively. We intended to grow small (<1 mm) single crystals because of the size limitations set by the cantilevers, therefore the growth time was limited to 3~4 days. Long rod-shaped crystals were selected for our measurement. Polycrystalline tellurium was grown by direct cooling of Tellurium vapor. Tellurium pieces were sealed in a quartz tube, heated above its boiling point (988 °C), and then furnace cooled to room temperature by switching off the box furnace. Black droplets with a metallic luster were found on the sidewalls of the quartz tube.

**Phonon-AM:**

Microscopically, the phonon-AM, $\boldsymbol{J}_{ph}$, arises from the circular motion of ionic atoms, which is defined as[4]:

$$\boldsymbol{J}_{ph} = \sum_{l\alpha} \boldsymbol{u}_{l\alpha} \times \dot{\boldsymbol{u}}_{l\alpha} \quad (M1)$$

where $\boldsymbol{u}_{l\alpha}$ represents the atomic displacement of atom α in the *l*th unit cell. In quantum mechanics, $\boldsymbol{u}_{l\alpha}$ can be written as:

$$\boldsymbol{u}_{l\alpha} = \sum_{\boldsymbol{k},\sigma} \boldsymbol{\epsilon}_\sigma(\boldsymbol{k}) e^{i(\boldsymbol{R}_l \cdot \boldsymbol{k} - \omega_\sigma(\boldsymbol{k})t)} \sqrt{\frac{\hbar}{2\omega_\sigma(\boldsymbol{k})N}} a_{\boldsymbol{k},\sigma} + H.c. \quad (M2)$$

Combining Eqs. (M1) and (M2), one obtains $\boldsymbol{J}_{ph} = \sum_{\boldsymbol{k},\sigma} \boldsymbol{l}_\sigma(\boldsymbol{k}) \left[ f_0(\omega_\sigma(\boldsymbol{k})) + \frac{1}{2} \right]$ and $\boldsymbol{l}_\sigma(\boldsymbol{k}) = \hbar \boldsymbol{\epsilon}_\sigma^\dagger(\boldsymbol{k}) \boldsymbol{M} \boldsymbol{\epsilon}_\sigma(\boldsymbol{k})$ in equilibrium. The matrix $\boldsymbol{M}$ is the tensor product of the unit matrix and the generator of *SO*(3) rotation for a unit cell with *N* atoms given by $(M_i)_{jk} = I_{N \times N} \otimes (-i)\varepsilon_{ijk}$ (*i, j, k* = *x, y, z*)[5]. At zero temperature, *T*=0, $\boldsymbol{J}_{ph} = \sum_{\boldsymbol{k},\sigma} \frac{1}{2} \boldsymbol{l}_\sigma(\boldsymbol{k})$.

**Torque measurements:**

The experiment is conducted using the Optical Multi-Function Probe (OMFP) with Raman Spectroscopy Option from Quantum Design. The probe is inserted into a DynaCool system (Quantum Design) for sample environment control. The OMFP insert contains a piezoelectric xyz-positioner for z-focusing and xy-moving. Once focused, the samples are moved relative to the laser spot for heating at different locations. The cantilevers are PRSA 300×100μm TL Probes from SCL-Sensor Tech. The cantilevers have a built-in Wheatstone bridge whose response changes with the cantilever bending. In the setup shown in Fig.2, the phonon-AM induced torques on each cantilever will show opposite signs. While a precise calibration is generally difficult for cantilever-based devices and techniques, we quantify our measurement by recording resistance readings before and after placing a ~0.15 mg Tellurium single crystal on the cantilever. Based on the torque generated by the weight of the sample, we obtain a scaling-coefficient in units of Nm/Ohm. Samples are anchored to the cantilever by N-grease, which solidifies at low temperatures.

**Laser power dependence of torque response:**

The excitation used in this experiment is a green laser ($\lambda = 532\ nm$). We tested the torque-laser power relation by gradually ramping up the laser power from 0 mW to 25 mW. The torque response on each cantilever has opposite signs and a linear dependence on the applied power (Figure M1).

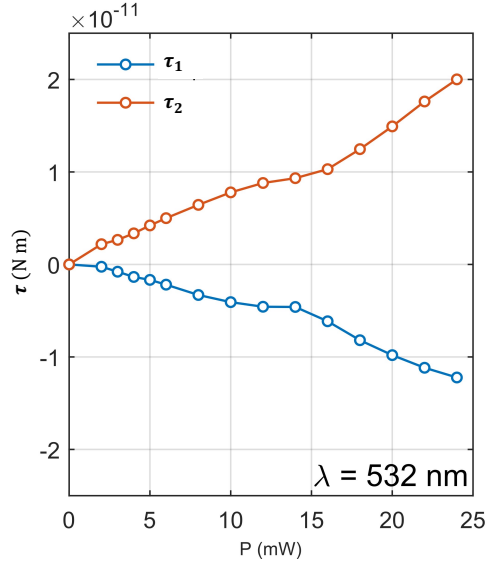

**Figure M1: Laser power (P) dependence of $\tau$.** Cantilevers 1 and 2 show opposite torque. The magnitude of $\Delta\tau$ approximately follows a linear dependence of laser power and consistent with the theoretical estimation of Eq (4), $\delta J_{ph} \propto \Delta T$.

**The laser heating effects on cantilevers and samples:**

The laser will not only generate thermal flows on samples, it will also induce heating effects, such as the thermal expansion, on both samples and cantilevers. This trivial heating effect show contributions to all our measurements. For instance, finite torque responses are observed in polycrystalline Te at low temperature and single crystal Te at the room temperature. The heating effect is also obvious in temperature and position dependent measurement of poly crystalline Te shown in Fig. 2**d** and Fig. 3**c**. In Fig. 2**d**, the $\tau$ is reduced at elevated temperature, where the heating effect is reduced. In Fig. 3**c**, larger $\tau$ is observed on the cantilever that is closer to the laser spot. In our measurements, we find that the heating effect is slightly larger on the polycrystal, which can be related to the differences in geometries, thermal conducting and thermal expansion behaviors between the poly and single crystals. It is important to note that the laser heating effect always yields $\tau$ on both cantilevers with the same sign, qualitatively different from phonon-AM induced $\tau$ in our experimental geometry.

**Acknowledgements**

We thank Andrew Christianson, Chenyun Hua, Raphael Hermann, Andrew May, Michael McGuire, Brian Sales, Ruixing Zhang, and Tiantian Zhang for stimulating discussions. This research was supported by the U.S. Department of Energy, Office of Science, Basic Energy Sciences, Materials Sciences and Engineering Division.

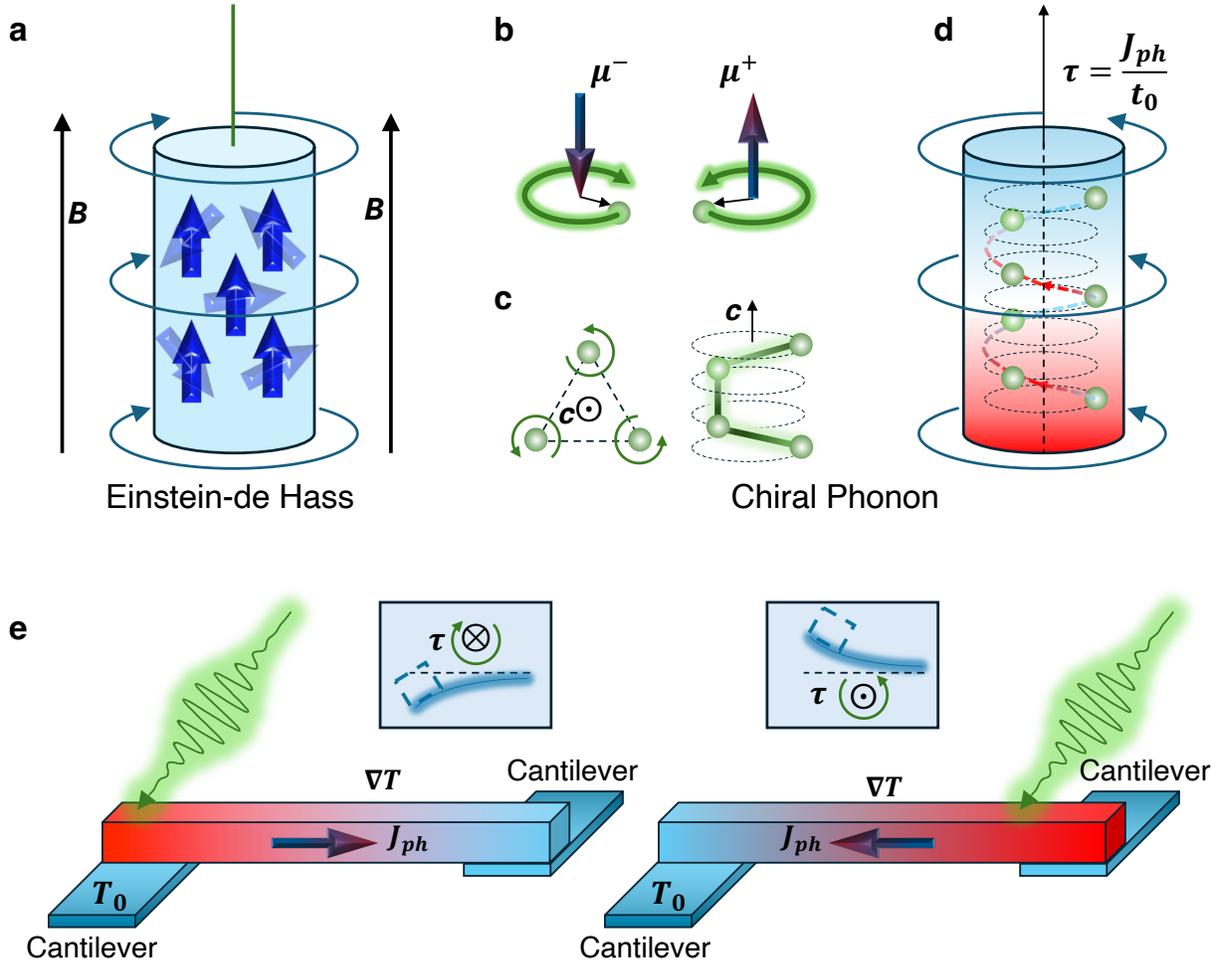

**Figure 1: The phonon AM induced mechanical rotation and torque. a**, The EdH effect that describes the mechanical rotation of a suspended ferromagnetic rod under external magnetic field. The mechanical rotation carries opposite AM to compensate the internal spin AM. **b**, circular motions of atoms that induce AM. **c**, Top and side view of one phonon mode, collective lattice motion, of atoms in Te. Right-handed motion propagating along the chiral c-axis gives rise to right-handed AM, $l_\sigma(k)$, of phonon mode $\sigma$. **d**, depicts the phonon EdH effect. The thermal gradient, $\nabla T$, breaks $\mathcal{T}$ and induces phonon-AM, $J_{ph}$. Due to AM conservation, the chiral crystal will mechanically rotate to compensate for the phonon-AM. With external constraints (cantilever contacts), the thermal gradient induced mechanical torque, $\tau = \frac{J_{ph}}{t_0}$, can be measured. **e**, schematics the cantilever-based device used here to determine the phonon-AM induced $\tau$. Samples (dashed blue boxes) are placed on top of cantilevers that are connected to cooling thermal reservoirs. A thermal gradient is introduced by shining a green light laser ($\lambda = 532\ nm$) on the samples. The phonon-AM induced $\tau$ bends the cantilevers shown in the inset.

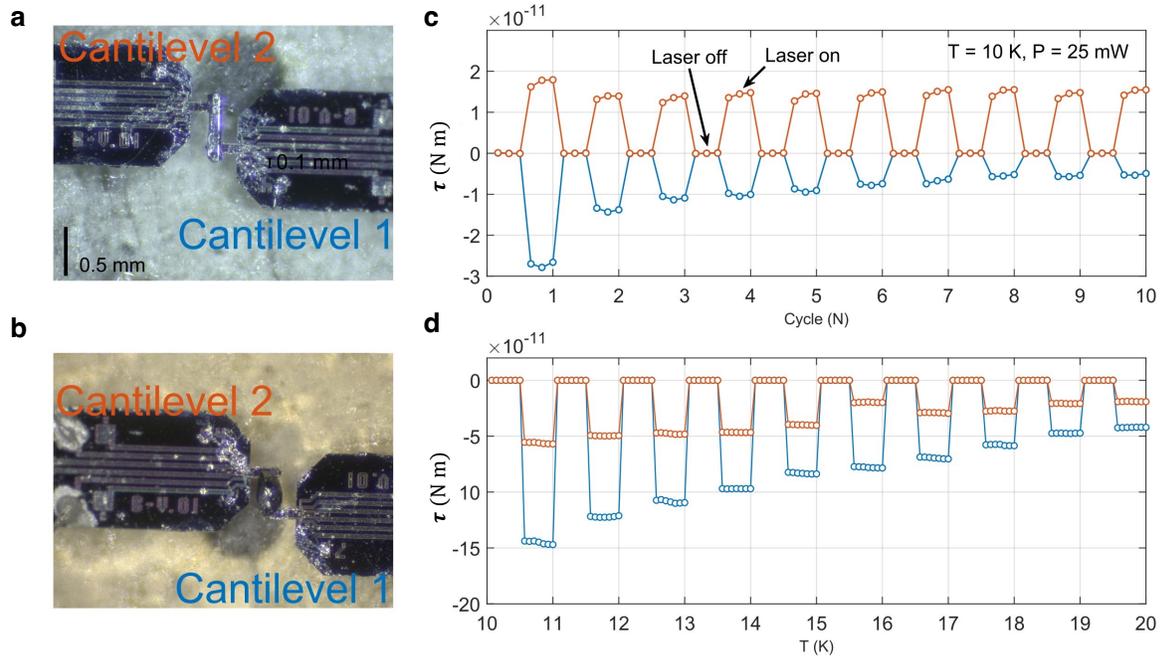

**Figure 2: Phonon-AM in chiral single crystal Te. a** and **b,** show photos of experimental setups for the single crystal and polycrystalline Te samples, respectively. The resistance readings from the right and left cantilevers are marked as Cantilever 1 and Cantilever 2, respectively. **c,** Torque responses for the single crystal Te sample. Switching the heating laser drives opposite $\tau$ from the two cantilevers over multiple cycles. **d,** Temperature dependent torque responses of the polycrystalline Te sample. The responses from the two cantilevers have the same sign when the laser is on. $\tau$ is reduced at elevated temperature, consistent with laser heating effect (see Methods). At a fixed temperature, the time interval between sequential data points in **c** and **d** is 5 s. When changing temperature (**d**), we turned off the laser and waited until the temperature is stabilized before we turn on the laser. We set the torque response of the device without laser heating at cycle, N=0, to be zero.

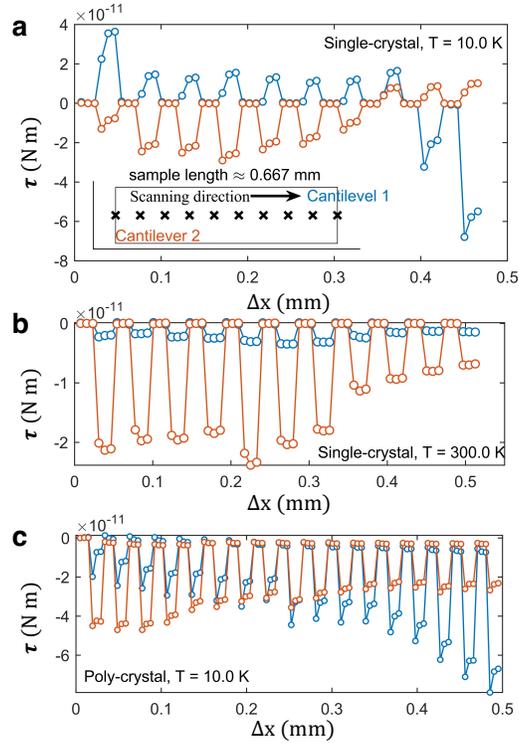

**Figure 3: Thermal and chirality control of phonon-AM. a,** Torque responses of the single crystal Te sample at 10 K. The laser moves from Cantilever 2 to Cantilever 1. The distance from the laser point to Cantilever 2 is marked as Δx. The torque responses of both Cantilever 1 and Cantilever 2 switch signs as the laser spot moves from Δx=0 to 0.5 mm. **b,** Torque responses of the single crystal Tellurium sample at 300 K. The sign of the torque responses on Cantilever 1 and Cantilever 2 remain unchanged as the laser spot moves. **c,** Torque responses of the polycrystalline Te sample at 10 K. The sign of the torque responses on Cantilever 1 and Cantilever 2 also remain unchanged as the laser spot is moved. The time interval between sequential data points in **a-c** is 5 s. We set the torque response of the device without laser heating to be zero.